# Critical Inertia Estimation for the Three U.S. Interconnections

Jiaojiao Dong[1], Fariha Hakim Sneha [1], Rocio Uria Martinez[2], Wen Wang[1], Yilu Liu[1],[2]
[1] The University of Tennessee, Knoxville, [2] Oak Ridge National Laboratory

***Abstract*—The rapid integration of inverter-based resources (IBRs) is reducing system inertia across U.S. power grids, raising concerns about frequency stability following large contingencies. This paper presents a simulation-based assessment of critical inertia—the minimum system inertia required to prevent first-stage under-frequency load shedding (UFLS) after the largest credible contingency—across the three major U.S. interconnections: Eastern Interconnection (EI), WECC, and ERCOT. Reduced-inertia scenarios are created by progressively replacing synchronous generators with IBRs, and dynamic simulations are performed using full-scale PSS®E and PowerWorld models. The results show that ERCOT reaches critical inertia at approximately 20% IBR penetration under the strict first-stage UFLS criterion—or approximately 58% if only the final UFLS stage must be avoided—compared with above 90% for EI and approximately 67%–68% for WECC. IBR shares in 2024 in the U.S. portions of EI, WECC, and ERCOT were 16%, 33%, and 44%, respectively, indicating varying proximity to critical inertia thresholds. These findings highlight the importance of full dynamic simulations to accurately estimate critical inertia and guide transmission planning under high renewable penetration scenarios.**

***Index Terms*--Critical inertia, inverter-based resources (IBRs), frequency stability, under-frequency load shedding (UFLS), ERCOT, WECC, Eastern Interconnection, power system simulation.**

## I. Introduction

The rapid increase of inverter-based resources (IBRs) in modern power systems has led to a sustained reduction in system inertia, raising growing concerns regarding frequency stability following large disturbances. System inertia plays a critical role in limiting the initial Rate of Change of Frequency (RoCoF) and providing sufficient time for frequency responsive resources (FRR) to arrest frequency decline following a contingency. As conventional synchronous generation is progressively displaced by IBRs with little or no inherent inertia, accurately quantifying the minimum inertia required for secure system operation has become an important challenge for transmission planners and reliability coordinators. Several major blackout events have highlighted the operational risks associated with insufficient system inertia, including the 2016 South Australia blackout, the 2019 Great Britain frequency event, and the 2021 Texas power crisis [1]–[3]. These incidents emphasize the importance of identifying the minimum inertia required to maintain secure operation under high renewable penetration conditions.

The North American Electric Reliability Corporation (NERC) defines critical inertia as the minimum level of system inertia necessary to ensure that frequency responsive resources have sufficient time to arrest frequency decline and prevent activation of the first stage of under-frequency load shedding (UFLS) following the largest credible contingency [4]. This definition directly links inertia adequacy to post-contingency frequency dynamics and protection thresholds. Consequently, critical inertia assessment methods must accurately capture dynamic frequency behavior and system-wide interactions under severe disturbances.

The broader impacts of declining inertia have been widely discussed in the literature [5]-[13]. These studies show that reduced inertia increases RoCoF, lowers nadir frequency, reduces frequency stability margins, and increases the need for faster frequency response. Low rotational inertia has been shown to significantly affect power system stability and operation [7]. Beyond frequency stability, high IBR penetration has also been associated with emerging oscillation phenomena, including ultra-low frequency and sub-synchronous oscillations, which pose additional challenges for system monitoring and stability assessment [8]. Reduced rotational inertia has also been associated with lower critical clearing time, resulting in faster transient dynamics and more stringent protection requirements under severe disturbances [9]. Comprehensive review studies have summarized inertia requirements, synthetic inertia techniques, fast frequency response strategies, and operational challenges associated with converter-dominated systems [10]-[12]. More recent studies have further investigated inertia and frequency security under high renewable penetration, considering virtual inertia and primary frequency response requirements [13], [14].

Several methods have been proposed in the literature to estimate critical inertia levels, often based on simplified frequency response models, analytical approximations, or reduced-order system representations. Existing works have also addressed related topics, including inertia estimation, inertia monitoring, frequency response analysis, and frequency support mechanisms. Recent work has further explored the feasibility of online inertia estimation by leveraging generator

This work was primarily supported by DOE/H2O office.

event signatures in large interconnection systems [15]. However, comparatively fewer studies directly investigate the minimum inertia required for secure system operation. Industry reports and technical studies have also examined the relationships among inertia level, RoCoF, frequency nadir, fast frequency response, and frequency containment requirements in low-inertia systems [16], [17].

One of the earliest practical discussions on critical inertia assessment was presented in the ERCOT whitepaper published in 2018 [18]. The study evaluated system frequency response following the largest credible contingency under varying inertia conditions. Reduced-inertia operating conditions were created by progressively replacing synchronous generators with inverter-based renewable generation. The critical inertia was identified as the inertia level at which system frequency reaches the 59.3 Hz UFLS threshold while still allowing sufficient response time from Load Resources.

Several subsequent studies and industry reports adopted similar simulation-based approaches for critical inertia assessment [19]-[22]. In these approaches, multiple operating conditions with different inertia levels are created, and the most credible contingency is simulated for each operating condition. Frequency response characteristics such as frequency nadir, UFLS activation, and RoCoF are then evaluated to determine the minimum inertia satisfying system security constraints. In UFLS-based approaches, critical inertia is defined as the minimum inertia level that avoids triggering under-frequency load shedding. In RoCoF-based approaches, critical inertia corresponds to the minimum inertia satisfying grid-code RoCoF limits. Since stricter RoCoF thresholds require slower frequency decay, lower RoCoF limits generally lead to higher critical inertia requirements. Although these methods are physically intuitive, they generally require extensive time-domain simulations under varying inertia conditions, making them computationally expensive for large-scale systems.

Recent studies have further investigated the relationship between renewable penetration and critical inertia thresholds using dynamic simulation and stability analysis techniques [23], [24]. Critical inertia thresholds under increasing renewable penetration were investigated in [23], where progressive replacement of synchronous generators with renewable resources was shown to significantly reduce system inertia and frequency stability margins. Minimum rotational inertia requirements based on system frequency dynamics and frequency deviation constraints were investigated in [24].

To reduce the computational burden associated with repeated simulations, analytical and reduced-order approaches have also been proposed [25], [26]. An energy-balance-based approach for critical inertia calculation was proposed in [25], where critical inertia was determined directly from imbalance energy following a disturbance instead of repeatedly simulating multiple inertia conditions. A linearized analytical method for critical inertia calculation in renewable-integrated power systems was later proposed in [26]. The method incorporated both frequency nadir and RoCoF constraints through a linearized representation of primary frequency response and derived analytical expressions for critical inertia using equivalent rotor motion equations.

Other recent studies have extended critical inertia assessment beyond conventional simulation and analytical methods by considering converter control, optimization, wind frequency regulation, uncertainty, and data-driven prediction [27]-[33]. For converter-dominated systems, the role of grid-forming control and storage-based virtual inertia was examined in [27], where a BESS-supported photovoltaic microgrid was used to determine the inertia level required to satisfy RoCoF and frequency-nadir constraints. Optimization-based formulations were also developed to estimate the minimum inertia needed to preserve transient stability in smart grids, with the objective of identifying vulnerable buses and supporting planning decisions [28].

More recent studies have incorporated frequency regulation capability and renewable uncertainty into inertia assessment. In [29], wind turbine frequency regulation was included through a multi-machine aggregation model, and minimum inertia was evaluated using both RoCoF and maximum frequency-deviation constraints. Data-driven methods have also been introduced to reduce the computational burden of repeated dynamic simulations. A deep-learning-based framework using an improved conservative convolutional neural network was proposed in [30] to predict frequency stability indicators and iteratively estimate critical inertia for renewable power systems. A probabilistic framework based on Copula-GAN scenario synthesis was further developed in [31] to evaluate critical inertia under correlated renewable generation, load uncertainty, and extreme operating scenarios.

In addition to direct inertia estimation, recent works have proposed broader frequency-security metrics and reserve co-optimization frameworks for low-inertia systems. A frequency stability margin index was developed in [32] to quantify the maximum tolerable power deficit while considering converter-based generation, equivalent inertia, RoCoF constraints, and frequency regulation capability. Similarly, a co-optimization framework for primary frequency response and fast frequency response reserves was proposed in [33], where linearized tri-criteria frequency constraints were used to satisfy RoCoF, frequency nadir, and quasi-steady-state frequency requirements.

Despite these developments, many existing approaches still rely heavily on repeated time-domain simulations under multiple inertia conditions, which become computationally expensive for large-scale interconnection studies. Furthermore, the applicability and robustness of existing methods for large, highly meshed interconnections remain insufficiently investigated. Most prior studies have primarily focused on ERCOT or smaller test systems, while comparatively limited attention has been given to full-scale assessments of the Eastern Interconnection (EI) and the Western Electricity Coordinating Council (WECC).

A preliminary comparative evaluation of multiple critical inertia calculation methods using the ERCOT system was previously presented in [34]. In that study, several methods incorporating load response and frequency support mechanisms were compared, resulting in lower critical inertia estimates. In contrast, the present work evaluates critical inertia based solely on the inherent frequency response of the system without relying on load response, UFLS assistance, or other emergency corrective actions. Consequently, the critical inertia values reported in this study are intentionally more conservative. The present work further extends beyond the preliminary investigation by providing broader interconnection-level analysis and additional evaluation of critical inertia characteristics under high IBR penetration.

This paper presents a simulation-based assessment of critical inertia for all three major U.S. interconnections—EI, WECC, and ERCOT—using detailed full-scale dynamic models implemented in PSS®E and PowerWorld. Unlike simplified analytical approaches, the proposed framework directly aligns with the NERC definition of critical inertia by explicitly evaluating whether UFLS activation is avoided following the largest credible contingency. Reduced-inertia operating conditions are systematically created by progressively replacing conventional synchronous generators with IBRs. This allows the evolution of frequency response characteristics to be evaluated under realistic transition pathways.

The major contributions of this paper are summarized as follows:

1. Development of a simulation-based critical inertia estimation framework closely aligned with the NERC definition and capable of avoiding optimistic bias associated with simplified analytical approaches.
2. Comprehensive evaluation of critical inertia for EI, WECC, and ERCOT using detailed full-interconnection dynamic models.
3. Quantification of both critical inertia levels and corresponding IBR penetration thresholds to provide insights for transmission planning and renewable integration studies.

The remainder of this paper is organized as follows. Section II describes the critical inertia assessment methodology and simulation setup. Section III discusses the simulation results for EI, WECC, and ERCOT. Finally, Section IV concludes the paper.

## II. Simulation Framework and Methodology

This section describes the simulation-based framework used to estimate critical inertia in large-scale power systems. The proposed methodology is designed to directly align with the NERC definition of critical inertia by explicitly evaluating post-contingency frequency performance and UFLS activation under progressively reduced inertia conditions.

### *A. Overview of the Critical Inertia Assessment Framework*

The proposed framework evaluates system frequency response following the largest credible contingency under a range of inertia levels. Rather than inferring critical inertia from analytical approximations or proxy metrics, critical inertia is determined through time-domain dynamic simulations that capture the interaction between inertia, frequency responsive resources, and UFLS protection.

The overall assessment procedure consists of the following steps: 1) Establish a base-case operating condition using a full interconnection dynamic model. 2) Identify the largest credible contingency for the system under study. 3) Systematically reduce system inertia by replacing synchronous generators with IBRs. 4) Perform dynamic simulations for each reduced-inertia condition. 5) Monitor system frequency and UFLS relay behavior. 6) Identify the minimum inertia level at which first-stage UFLS activation is avoided. This process is applied independently to the EI, WECC, and ERCOT systems using consistent assumptions and evaluation criteria.

### *B. Creation of Reduced-Inertia Operating Conditions*

Reduced-inertia conditions are created by gradually replacing conventional synchronous generating units with IBRs. This approach reflects plausible generation transition pathways and preserves the structural characteristics of the original system models.

At each replacement step, synchronous generators are retired or converted to IBR equivalents with zero or negligible inertia contribution. System load and total generation are maintained to ensure power balance. Unit commitment and dispatch are adjusted to reflect the new resource mix while preserving credible operating conditions.

The total system inertia is calculated as the sum of inertia contributions from all remaining synchronous machines. By progressively increasing IBR penetration, a wide range of inertia levels is evaluated, enabling identification of the inertia threshold at which frequency performance becomes inadequate. This methodology avoids artificially scaling inertia parameters or using equivalent system inertia constants, both of which may mask localized inertia depletion and spatially dependent frequency behavior.

### *C. Dynamic Simulation Setup*

Studies are conducted using PSS®E with full interconnection dynamic models for EI and ERCOT. Studies in WECC are evaluated using a full dynamic model implemented in PowerWorld. Time-domain simulations are performed following the initiation of the largest credible contingency, typically a major generation loss.

Key elements of the simulation setup include: 1) Detailed generator, governor, and turbine models for synchronous units. 2) Explicit modeling of frequency responsive resources through governor action and primary frequency control. 3) Representation of UFLS relays and associated frequency thresholds. 4) Consistent simulation time steps and numerical integration settings across all cases. System frequency is monitored at multiple buses to capture interconnection-wide frequency behavior. The minimum observed system frequency is compared against UFLS thresholds to determine whether first-stage load shedding is activated.

### *D. Contingency Selection and Evaluation Criteria*

For each interconnection, the largest credible contingency is identified based on system operating conditions and reliability criteria. This contingency typically corresponds to the loss of the largest online generating unit or generation block.

The evaluation criterion for critical inertia is binary and operationally grounded. Secure operation is defined as the case in which first-stage UFLS is not activated following the contingency, whereas insecure operation is defined as the case in which first-stage UFLS is activated at any point during the simulation. Critical inertia is defined as the minimum system inertia level that maintains secure operation under the largest credible contingency. This definition directly reflects the NERC criterion and avoids reliance on surrogate metrics, such as RoCoF limits or analytically estimated frequency nadirs. In this study, critical inertia is evaluated without relying on load shedding, load response, virtual inertia from IBRs, or other emergency corrective actions to arrest frequency decline. The objective is to identify the minimum inertia required for the system to remain secure following the largest credible contingency using inherent system frequency response alone. Consequently, the resulting critical inertia values may be more conservative than estimates obtained from methodologies that explicitly account for load response or UFLS assistance.

### *E. Implementation Across U.S. Interconnections*

The proposed framework is applied consistently to the EI, WECC, and ERCOT systems. While the same methodology is used for all three interconnections, system-specific characteristics—including size, topology, generation mix, and UFLS settings—are preserved in each case. By applying a unified simulation-based methodology to all three U.S. power grids, the resulting critical inertia estimates enable direct comparison across interconnections and provide insight into how system size and structure influence inertia requirements.

Although this study focuses on EI, WECC, and ERCOT, the proposed framework is generalizable to any power system with a validated dynamic model, defined credible contingency criteria, and known UFLS settings. For another system, the same procedure can be applied by identifying the largest credible contingency, creating reduced-inertia operating conditions through resource replacement scenarios, performing time-domain simulations, and determining the minimum inertia level that avoids first-stage UFLS activation. Therefore, the framework can support inertia adequacy assessment for both interconnection-scale systems and regional grids undergoing high IBR integration.

## III. CASE STUDY

This section presents the system models, study scenarios, and critical inertia assessment results for the three major U.S. interconnections: EI, WECC, and ERCOT. For each interconnection, a full dynamic model is used to evaluate frequency response under progressively reduced inertia conditions following the largest credible contingency. Critical inertia is determined based on avoidance of first-stage UFLS activation, consistent with the methodology described in Section II.

### *A. ERCOT*

Fig. 1 shows the system frequency trajectories following the contingency at IBR penetration levels of 20%, 40%, 60%, and 80%. Three horizontal lines indicate the first three stages of UFLS at 59.3 Hz, 58.9 Hz, and 58.5 Hz, respectively. As IBR penetration increases, the initial rate of frequency decline becomes steeper and the frequency nadir decreases. At lower IBR penetration levels, the system frequency remains above the first-stage UFLS threshold, whereas higher IBR penetration results in UFLS activation due to insufficient time for responsive reserve service (RRS) to arrest the frequency decline.

Figure 2 summarizes the relationship between frequency nadir and IBR penetration level. Red markers indicate simulated data points obtained from dynamic simulations, while the blue curve represents an interpolated trend connecting these points. The results show a nonlinear degradation of frequency nadir with increasing IBR penetration. The critical inertia point is identified at the IBR penetration level where the interpolated frequency nadir intersects the first-stage UFLS threshold of 59.3 Hz. This intersection defines the minimum inertia level required to prevent UFLS activation in ERCOT under the studied operating condition. The estimated critical inertia level for ERCOT is 504 GVAs and would be reached after the simulated contingency for an IBR level of ~20%, as marked by the yellow star in Fig. 2. Also, if we only need to prevent the final stage UFLS activation with a threshold of 58.5 Hz, the estimated critical inertia level for ERCOT is 265 GVAs and would be reached after the simulated contingency for an IBR level of ~58%.

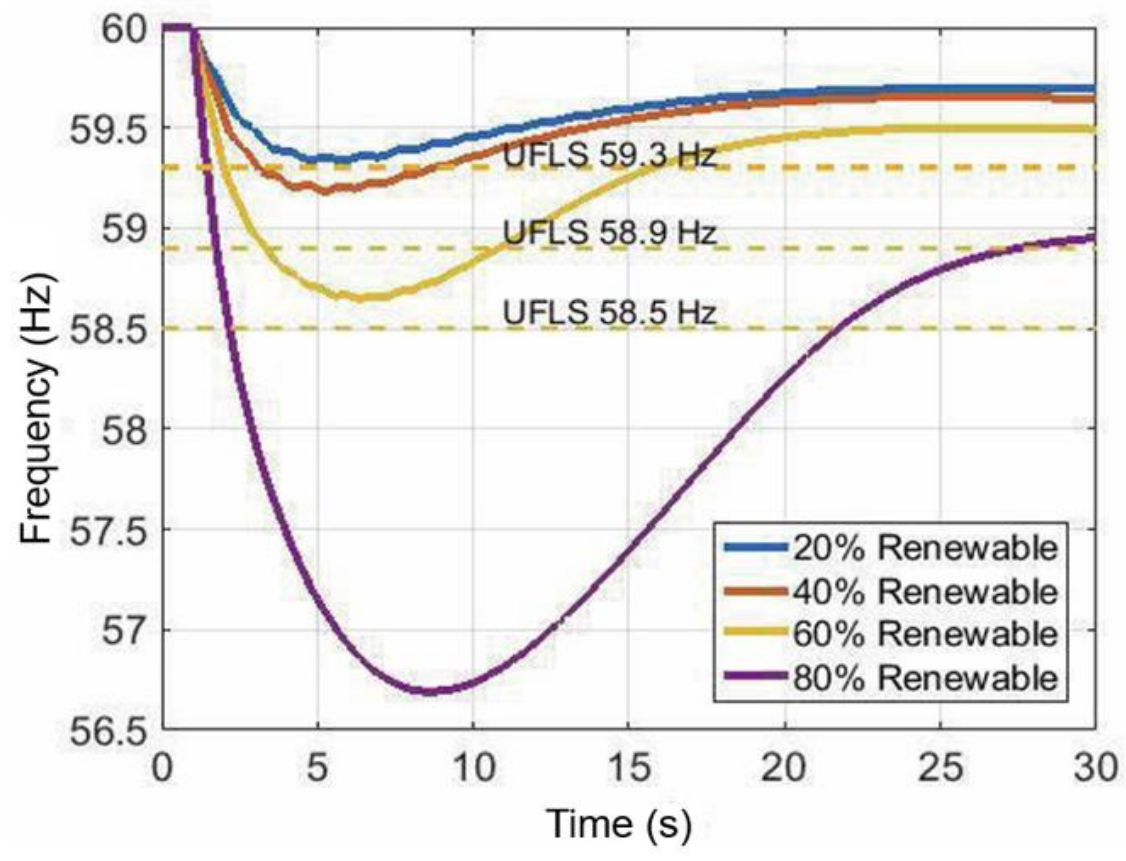

Figure 1. ERCOT System Frequency Response Following Largest Credible Contingency at Various IBR Penetration Levels

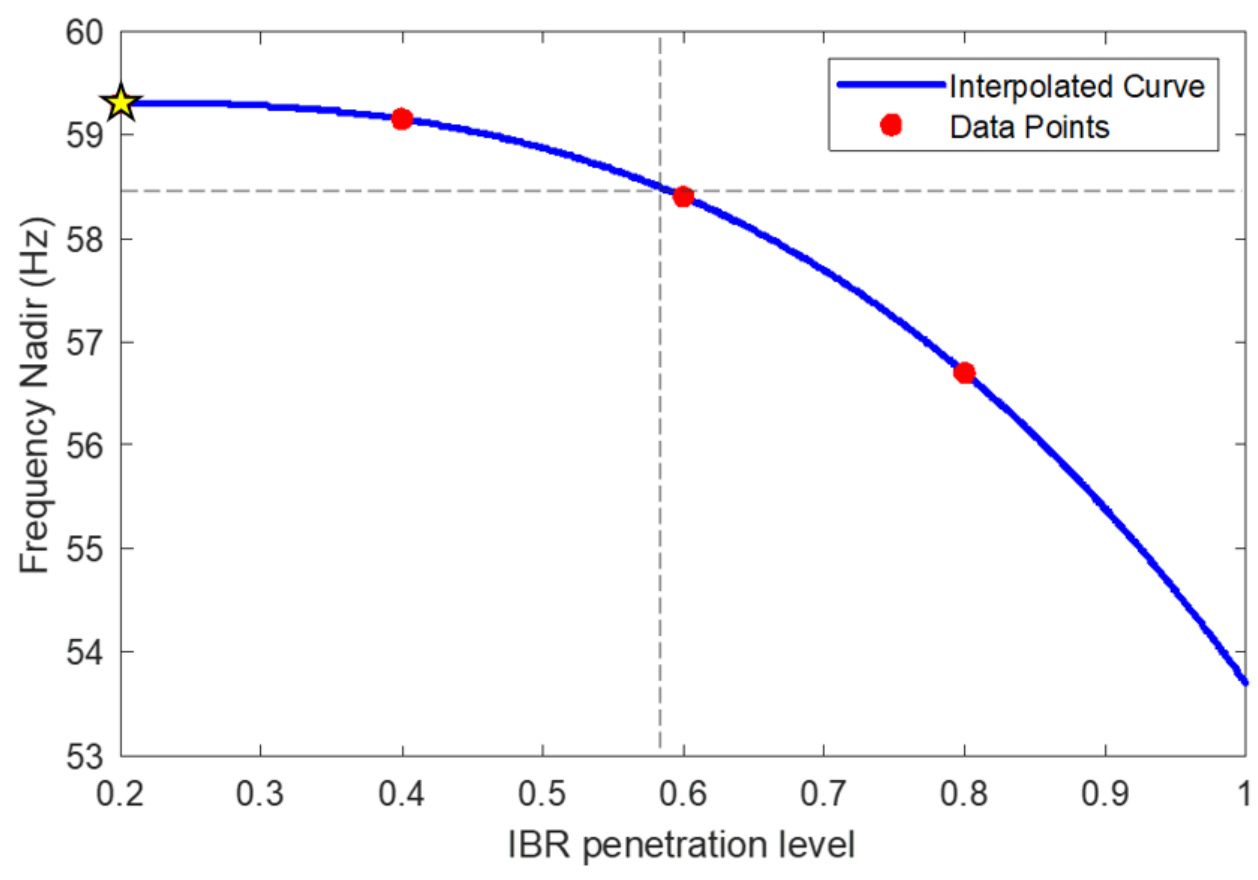

Figure 2. Relationship between Frequency Nadir and IBR Penetration Level in ERCOT

## *B. WECC*

The WECC system is evaluated under both base-case and reduced-inertia operating conditions to examine how increasing IBR penetration affects frequency response. The base-case result provides a reference for comparison, while the reduced-inertia cases illustrate the progressive degradation of frequency nadir as synchronous inertia is displaced. Fig. 3 shows the frequency response in the base-case scenario with 10% IBR penetration. Fig. 4 shows the frequency response under a reduced-inertia scenario with 60% IBR penetration. Compared with the base case, the reduced-inertia case exhibits a steeper initial frequency decline and a lower frequency nadir, approaching the first-stage UFLS activation threshold. Fig. 5 presents the relationship between frequency nadir and IBR penetration level. Red markers indicate simulated cases, and the blue curve is an interpolation connecting the points. The critical inertia is identified where the frequency nadir intersects the UFLS threshold (59.5 Hz), representing the minimum inertia required to prevent first-stage UFLS activation.

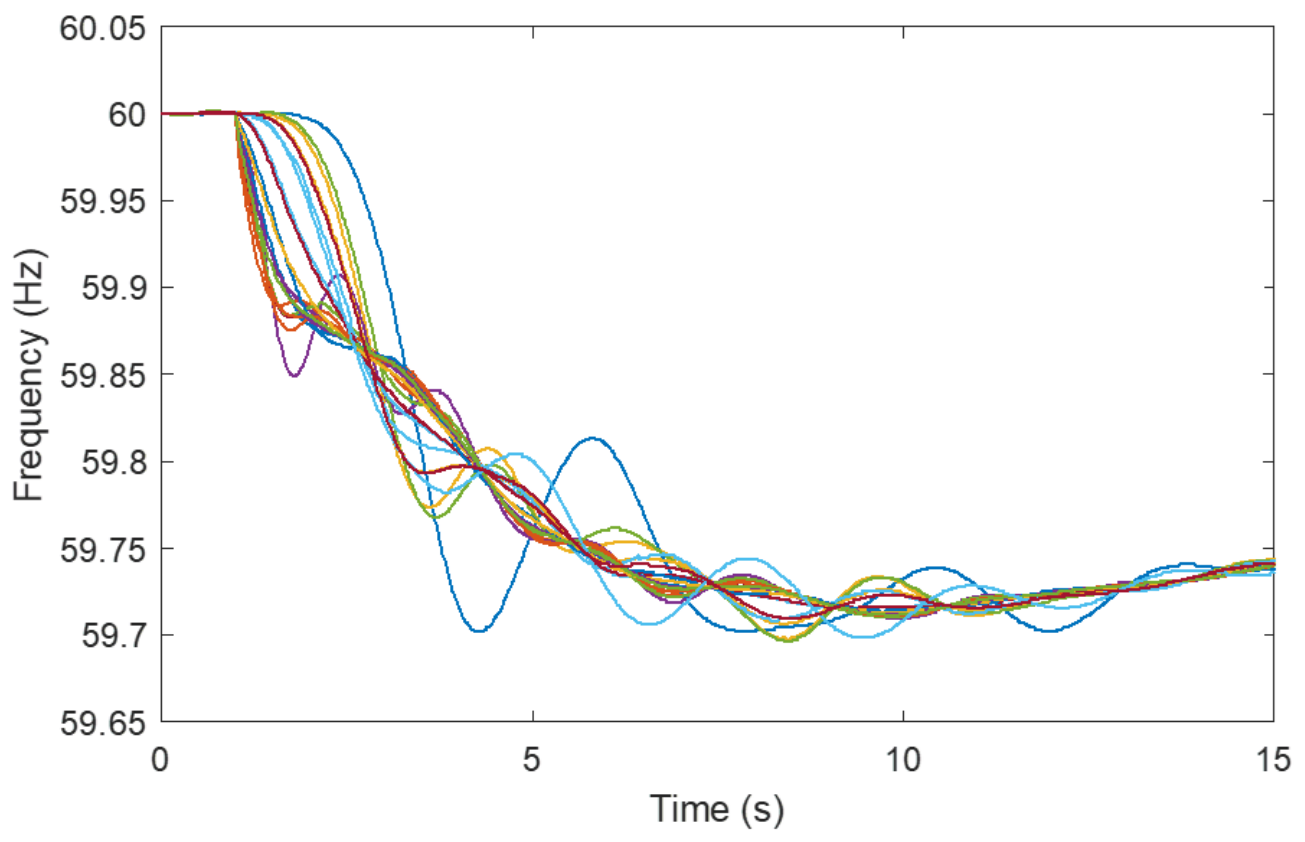

Figure 3. WECC Base-Case Frequency Response (10% IBR Penetration)

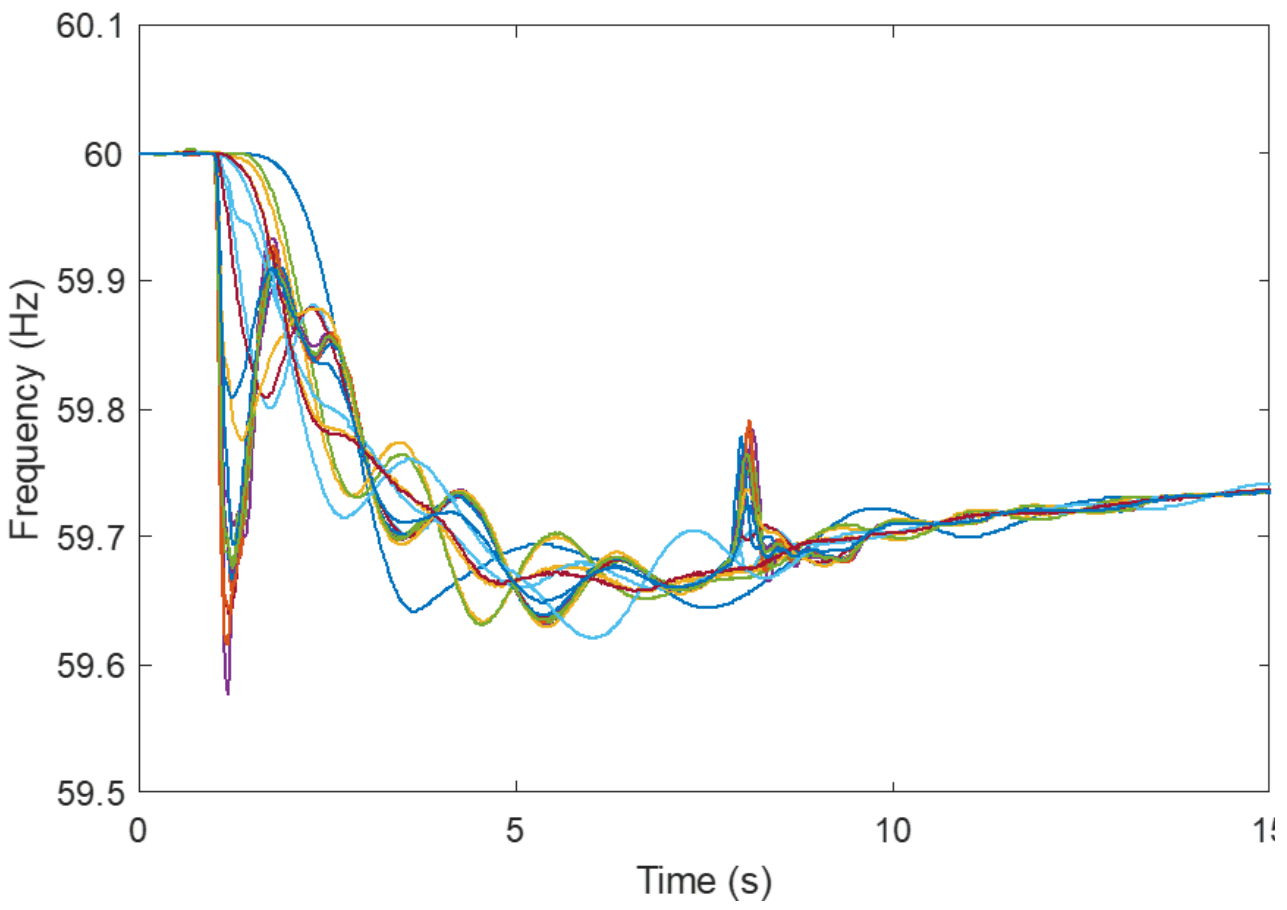

Figure 4. WECC Reduced-Inertia Frequency Response (60% IBR Penetration)

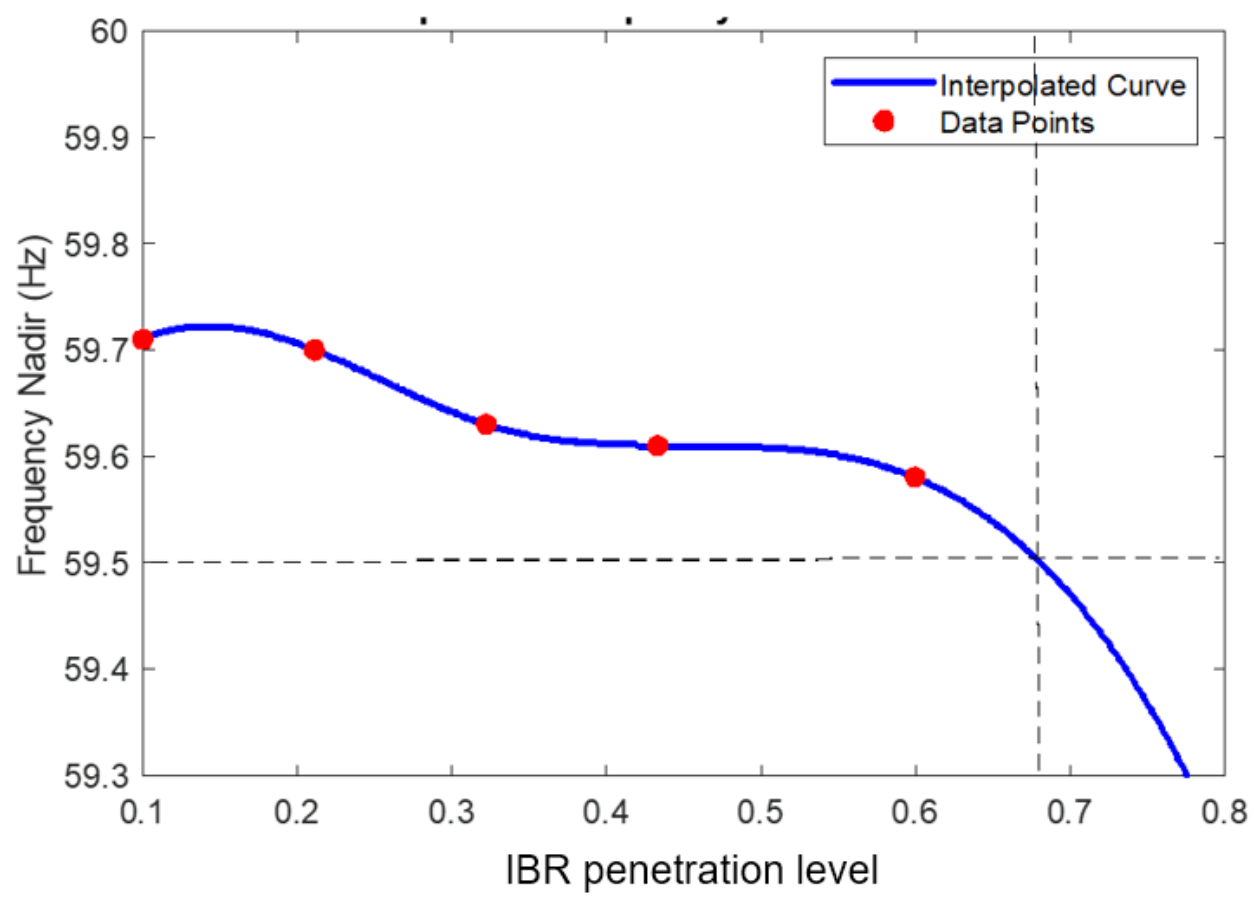

Figure 5. Relationship between Frequency Nadir and IBR Penetration Level in WECC with PowerWorld Model

WECC dynamic models with different IBR levels created in PSLF (Positive Sequence Load Flow) for a different study in our group produced a very similar estimate of critical inertia. Fig. 6 shows frequency trajectories for the base-case and 20%, 40%, 60%, and 80% IBR penetration levels, with a horizontal line indicating the UFLS threshold at 59.5 Hz. As IBR penetration increases, the frequency nadir decreases, and the system approaches UFLS activation. The PSLF-based results show a similar trend, supporting the robustness of the estimated WECC critical inertia threshold.

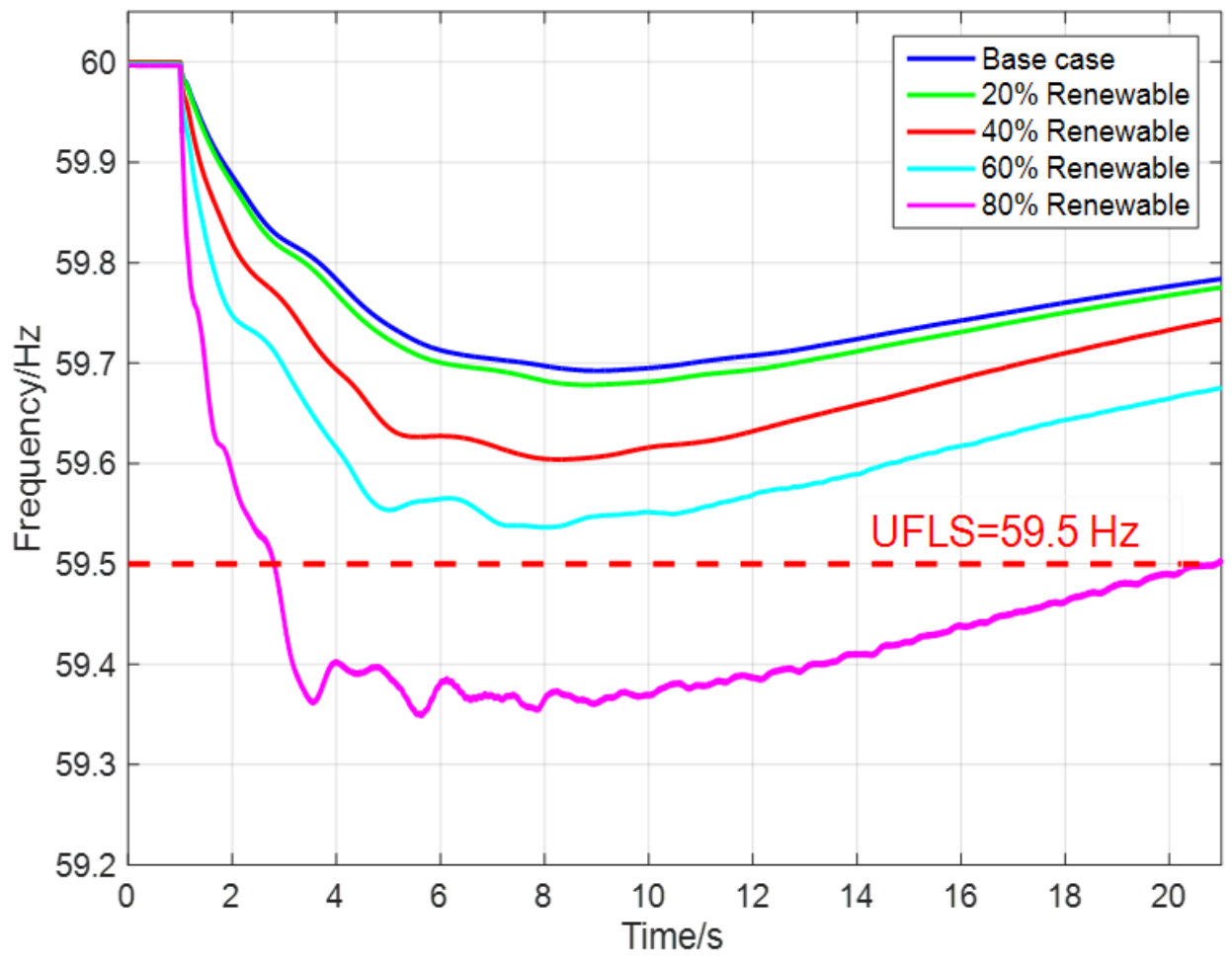

Figure 6. WECC Frequency Trajectories at Various IBR Penetration Levels (PSLF Model), with UFLS Threshold at 59.5 Hz

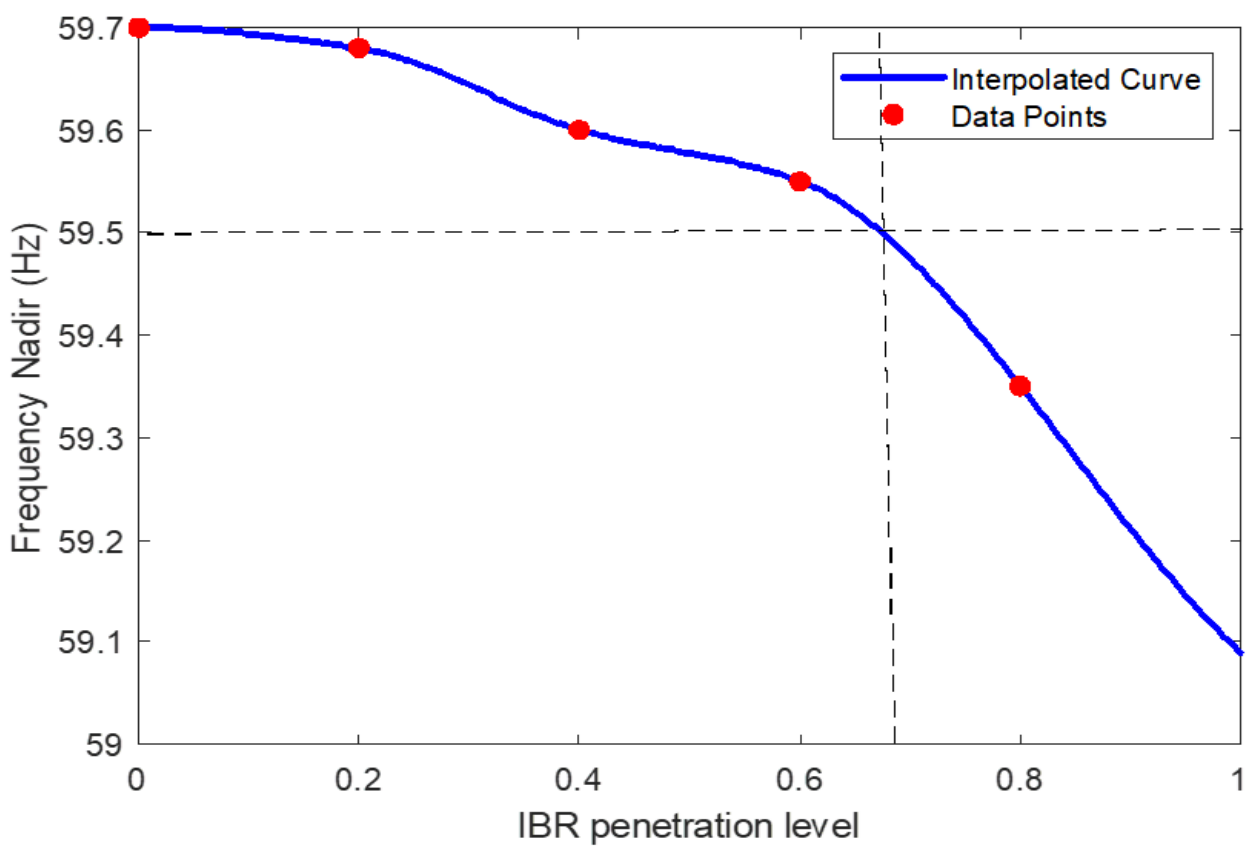

Figure 7. Relationship between Frequency Nadir and IBR Penetration Level in WECC using PSLF model

Fig. 7 summarizes the frequency nadir as a function of IBR penetration. Red markers correspond to simulated data points, and the blue curve represents the interpolated trend. The intersection of the curve with the UFLS threshold identifies the critical IBR penetration level (approximately 67–68%). Finally, Fig. 8 illustrates the inertia deficit (GVA·s), defined as the additional inertia required under current system conditions to maintain secure operation, as a function of increasing renewable energy penetration. The results highlight that higher renewable penetration increases the system's inertia deficit, emphasizing the need for careful planning to maintain frequency security in WECC.

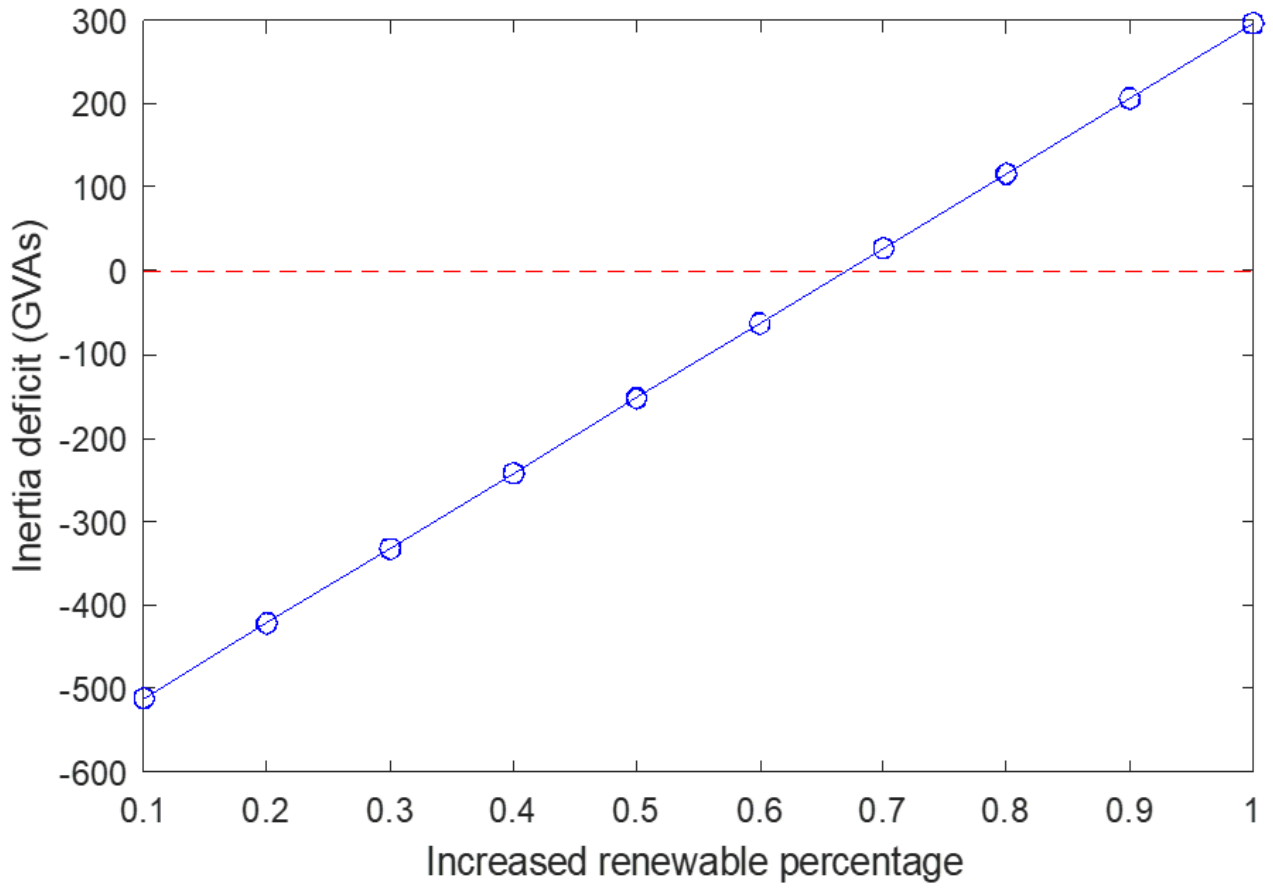

Figure 8. WECC Inertia Deficit vs. IBR Penetration Level

## C. EI

Critical inertia in the Eastern Interconnection (EI) is evaluated using a full dynamic model of the interconnection. Due to the large size and strong internal coupling of EI, the system is highly stable, making it difficult to reach the UFLS threshold of 59.5 Hz under typical contingencies. Therefore, marginally stable scenarios are used to identify critical inertia.

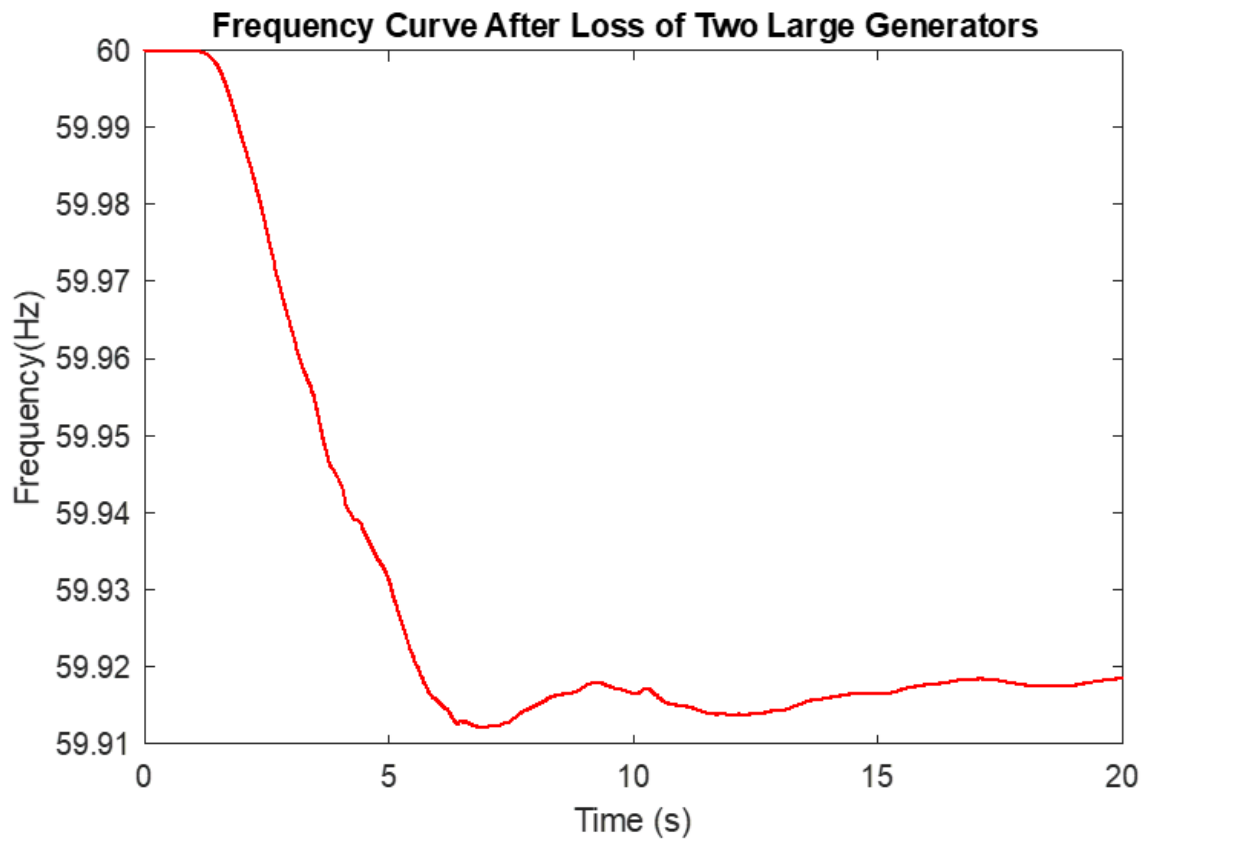

Figure 9. EI Base-Case System Frequency Response Following Loss of Two Large Generators

Fig. 9 shows the system frequency response following the loss of two large generators in the base-case EI scenario. A single curve is shown because the mean and median frequencies across different grid locations are nearly identical, reflecting the strong interconnection characteristics of EI. Figs. 10 and 11 show frequency trajectories under marginally stable and marginally unstable conditions, respectively. These cases are used to determine the minimum inertia required to maintain secure operation without triggering UFLS. The marginal cases are generated by progressively reducing synchronous generation until the post-contingency frequency response approaches the UFLS threshold.

Fig. 12 presents the relationship between frequency nadir and total system inertia (GVA·s). Red markers indicate simulated cases, and the blue curve represents the interpolated trend connecting the points. The critical inertia is identified at the inertia level (387 GVA·s, corresponding to >90% IBR penetration) where the frequency nadir approaches the marginally stable threshold, e.g., 59.5 Hz, providing a practical estimate for operational planning. Fig. 13 shows the inertia deficit versus increasing renewable penetration. The inertia deficit represents the additional inertia required under current system conditions to maintain marginal stability. As renewable penetration increases, the system inertia deficit grows, highlighting the need for careful resource and transmission planning to ensure frequency security in EI.

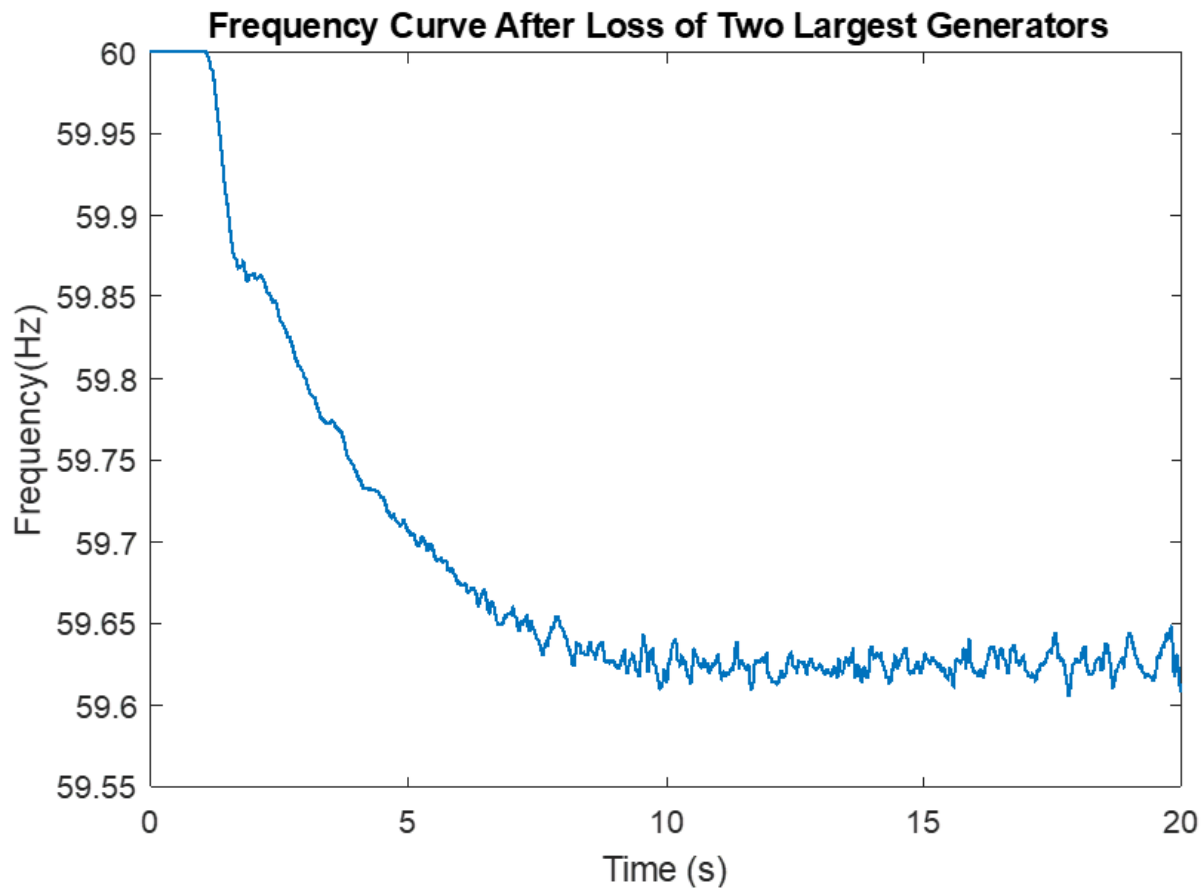

Figure 10. EI Frequency Trajectory Under Marginally Stable Condition

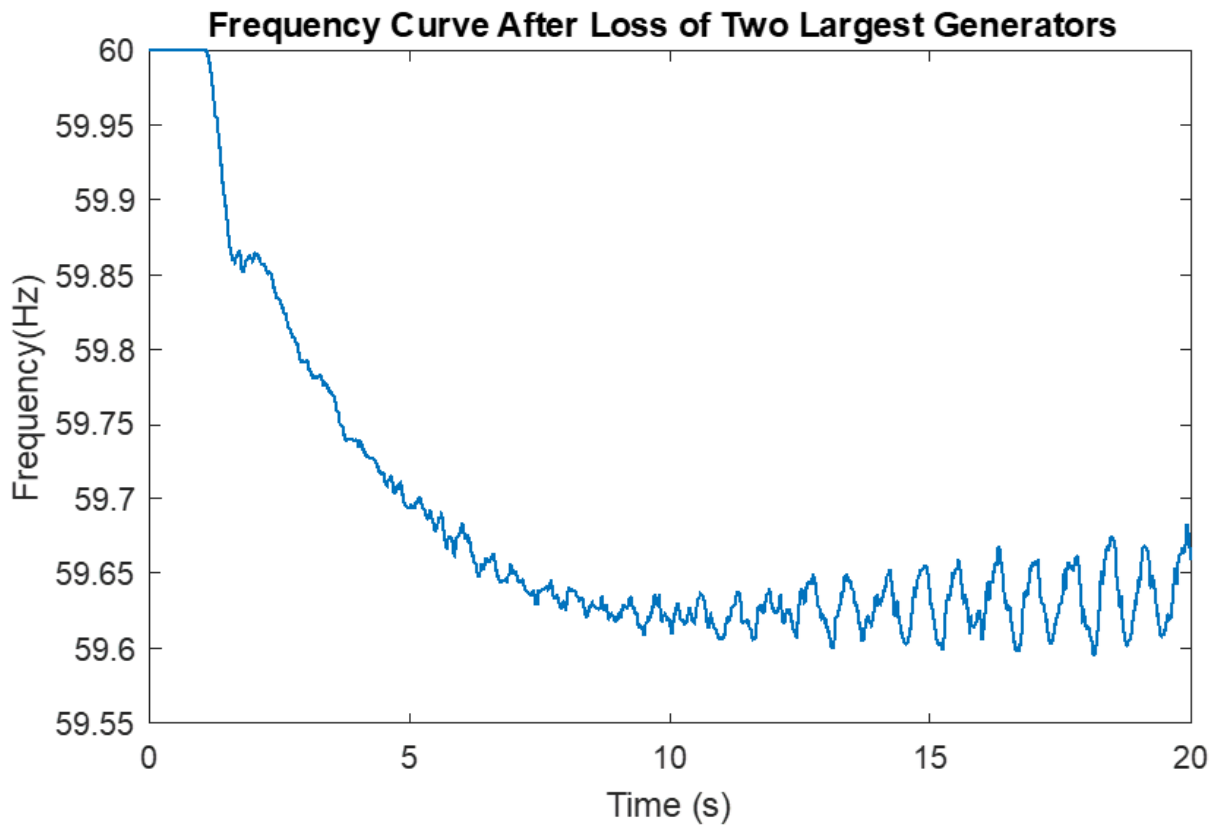

Figure 11. EI Frequency Trajectory Under Marginally Unstable Condition

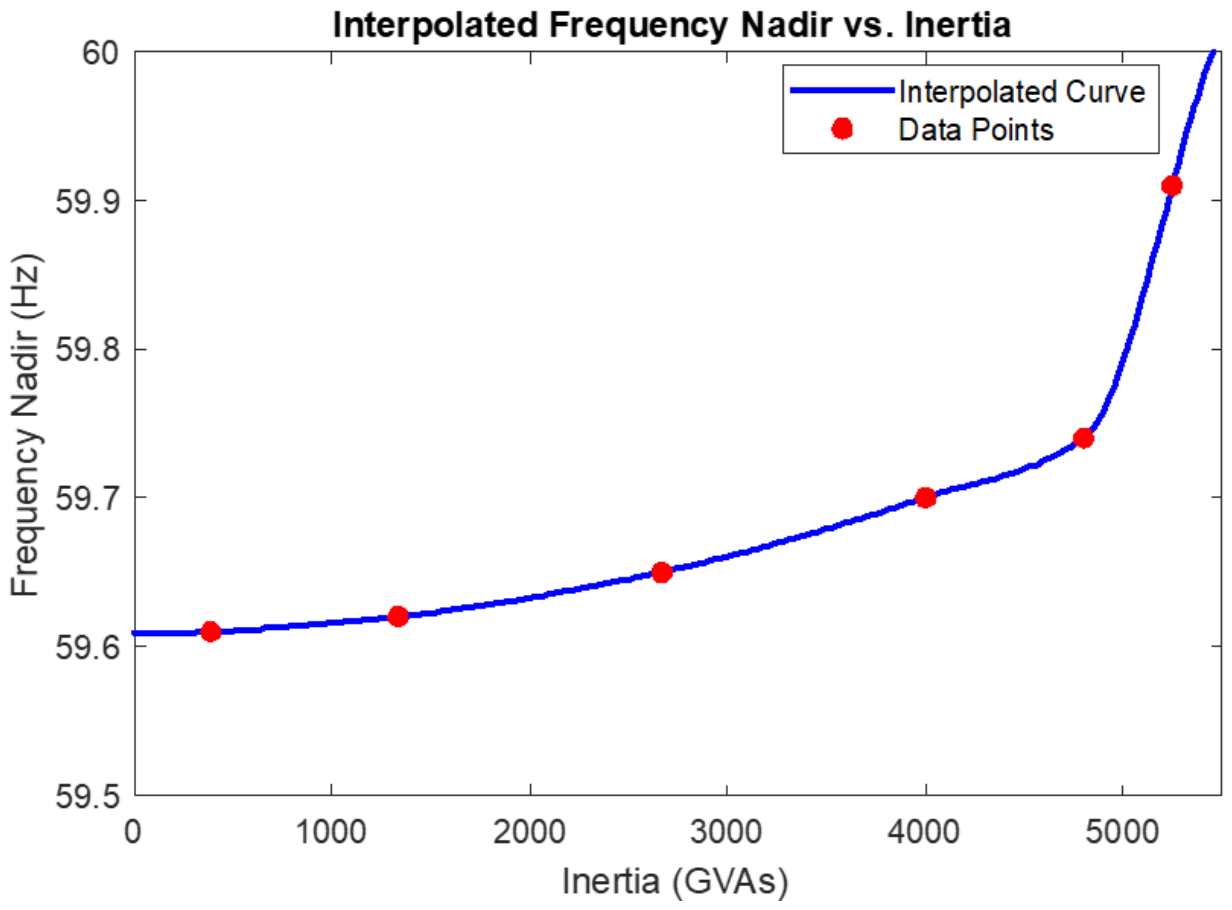

Figure 12. Relationship between Frequency Nadir and Total System Inertia in EI

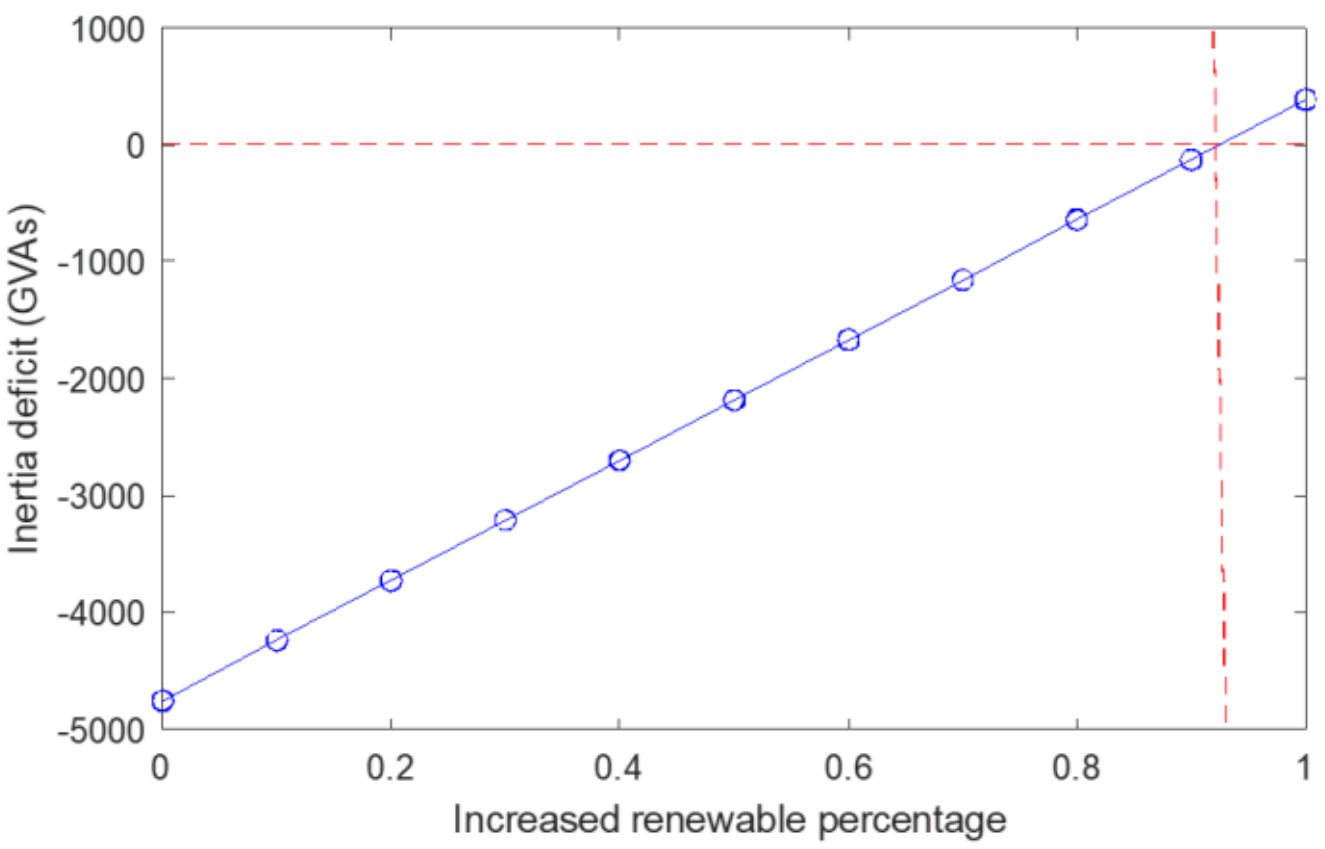

Figure 13. EI Inertia Deficit vs. IBR Penetration

### D. *Summary*

Table I summarizes the critical inertia assessment results for EI, WECC, and ERCOT. The table includes total system inertia, simulated contingencies, estimated critical inertia, IBR penetration level at which critical inertia is reached, and current IBR share based on 2024 EIA Form 860 data [35].

TABLE I. CRITICAL INERTIA ASSESSMENT SUMMARY FOR EI, WECC, AND ERCOT

| **Inter-connection** | EI | WECC | ERCOT |
|---|---|---|---|
| **Total inertia** | 5,250 GVAs | 808 GVAs | 443 GVAs |
| **Simulated contingency** | Loss of two large generators (combined capacity of 4,762 MW) | Loss of two large generators (combined capacity of 2,740 MW) | Loss of two largest generators (combined capacity of 2,750 MW) |
| **Estimated critical inertia** | 387 GVAs | 287–296 GVAs | 504 GVAs (no UFLS triggered) 265 GVAs (final UFLS stage avoided) |
| **IBR penetration at which critical inertia is reached** | >90% | 67%–68% | ~20% (no UFLS triggered) ~58% (final UFLS stage avoided) |
| **IBR share in 2024 in U.S. portions of the interconnections [33]** | 16% | 33% | 44% |

The comparison shows significant differences among the three interconnections. ERCOT reaches critical inertia at relatively moderate IBR penetration (~20%) with no USLF and (~58%) with final UFLS stage avoided, while EI, due to its large total inertia, reaches critical inertia only at very high IBR penetration (>90%). WECC, despite being a large interconnection, reaches critical inertia at 67–68% IBR penetration under marginally stable conditions. These results highlight both the scale and structural differences of the U.S. power grids and emphasize the importance of system-specific assessment of critical inertia for operational planning and integration of IBRs.

## IV. CONCLUSION

This study presents a comprehensive, simulation-based assessment of critical inertia for the three major U.S. interconnections—EI, WECC, and ERCOT. By replacing synchronous generation with IBRs and evaluating system frequency response under the largest credible contingencies, the minimum inertia levels required to prevent first-stage UFLS activation are identified. Key findings include:

- EI, due to its large total inertia, reaches critical inertia only at very high IBR penetration (>90%).
- WECC reaches critical inertia at 287–296 GVA·s, corresponding to 67–68% IBR penetration under marginally stable conditions.
- ERCOT reaches critical inertia at approximately 504 GVA·s (~20% IBR penetration) under the strict first-stage UFLS criterion, or approximately 265 GVA·s (~58% IBR penetration) if only the final UFLS stage must be avoided.

- Current IBR shares indicate that ERCOT is closest to its critical inertia limit, whereas WECC and EI remain farther from their respective thresholds.

These results emphasize the importance of system-specific, large-scale dynamic simulations for accurately assessing inertia needs in power grids with high renewable penetration. The proposed methodology provides actionable insights for transmission planning, operational decision-making, and integration of IBRs, ensuring frequency security under evolving generation mixes. The framework can also support future studies involving synthetic inertia, grid-forming IBRs, fast frequency response, and alternative UFLS settings.